\documentclass[%
 reprint,
 amsmath,amssymb,
 aps,
]{revtex4-2}

\usepackage{graphicx,float, placeins,xcolor}
\usepackage{subcaption}
\usepackage{dcolumn}
\usepackage{bm}
\begin{document}

\preprint{APS/123-QED}

\title{Topological Phase Diagram of Generalized SSH Models with Interactions}

\author{Yuxiao Hang\textsuperscript{1}}
\email{yhang@usc.edu}
\author{Stephan Haas\textsuperscript{1}}%
 \email{shaas@usc.edu}
 
\affiliation{\textsuperscript{1}Department of Physics and Astronomy, University of Southern California,
Los Angeles, CA 90089-0484, USA
}%

\date{\today}

\begin{abstract}
We investigate interacting Su-Schrieffer-Heeger (SSH) chains  with two- and three-site unit cells using density matrix renormalization group (DMRG) simulations. By selecting appropriate filling fractions and sweeping across interaction strength \( J_z \) and dimerization \( \delta \), we map out their phase diagrams and identify transition lines via entanglement entropy and magnetization measurements. In the two-site model, we observe the emergence of an interaction-induced antiferromagnetic intermediate phase between the topologically trivial and non-trivial regimes, as well as a critical region at negative \( J_z \) with suppressed magnetization and finite-size scaling of entanglement entropy. In contrast, the three-site model lacks an intermediate phase and exhibits asymmetric edge localization and antiferromagnetic ordering in both positive and negative \( J_z \) regimes.
We further examine the response of edge states to  Ising perturbations. In the two-site model, zero-energy edge modes are topologically protected and remain robust up to a finite interaction strength. However, in the three-site model, where the edge states reside at finite energy, this protection breaks down. Despite this, the edge-localized nature of these states survives in the form of polarized modes whose spatial profiles reflect the non-interacting limit.

\end{abstract}

\maketitle


\section{Introduction}
Since the discovery of topological insulators\cite{PhysRevLett.95.226801,doi:10.1126/science.1148047,RevModPhys.82.3045,RevModPhys.83.1057}, the study of quantum phase transitions in topological systems has attracted widespread interest. These transitions are characterized by changes in global topological invariants, such as the Zak phase, and the emergence or disappearance of symmetry-protected edge states. While non-interacting models have provided a foundational understanding—facilitated by topological band theory and symmetry analysis—the physics of interacting topological quantum systems remains more elusive.

This underdeveloped understanding is exemplified by effects which cannot be captured by single-particle approaches, such as interaction-induced suppression of topological phases and quantum-to-classical crossovers. Studying these effects requires access to inherently many-body quantities, such as entanglement entropy, which can only be obtained using advanced  techniques such as the Density Matrix Renormalization Group (DMRG).

Among various entanglement measures, the von Neumann entanglement entropy (EE) has proven especially powerful in characterizing quantum phases and detecting critical behavior\cite{Sachdev_1999,PhysRevLett.90.227902,PhysRevLett.101.010504,PhysRevB.81.064439}. Defined for a bipartition of the system into subsystems A and B, the EE is given by 
\begin{equation}
    S=-Tr(\rho_A\ln{\rho_A}),
\end{equation}
where $\rho_A$ is the reduced density matrix of the subsystem A.
In quantum systems at zero temperature, this quantity reflects long-range quantum correlations and can be used to distinguish gapped and gapless phases. In particular, in 1D conformal field theory predicts a logarithmic scaling of entanglement entropy at criticality, with different scaling behaviors signaling different universality classes of the phase transition.

Topological edge modes, protected by symmetries such as chiral or inversion symmetry, serve as another hallmark of non-trivial phases. In systems with open boundaries and gapped bulk spectra, these localized edge states arise as a consequence of the bulk-boundary correspondence\cite{ShunQingshen2012}. Their presence and robustness have been widely studied, both theoretically and experimentally\cite{PhysRevB.101.235155,PhysRevA.99.013833}, and their behavior under various filling conditions provides insight into the ground-state properties of topological phases.

Bond-alternating spin chains, such as the Su-Schrieffer-Heeger (SSH) model and its interacting generalizations, have played a central role in simulating one-dimensional topological insulators. The two-site unit cell SSH model with XXZ-type interactions is known to host symmetry-protected topological phases\cite{Qiang_2013,PhysRevB.24.5229}, and has been studied analytically\cite{PhysRevB.24.5229,PhysRevB.110.165145} and numerically\cite{PhysRevB.64.024410,Tzeng2016}. However, only limited work has so far been done to explore their entanglement structure or construct detailed phase diagrams in the presence of interactions. Even less is known about interacting SSH models with larger unit cells, such as the three-site unit cell bond-alternating XXZ chains studied here. Although the non-interacting version of this model preserves edge modes despite broken inversion symmetry\cite{PhysRevB.106.085109}, the effects of interactions on its entanglement properties and phase structure remain largely unexplored.

In this work, we address these open issues by studying the entanglement entropy and local observables in interacting SSH-type models with two- and three-site unit cells. Using DMRG, we map out their phase diagrams, identify distinct topological and trivial phases, and characterize the nature of the quantum phase transitions. Our results reveal interaction-driven modifications to topological regimes and uncover a quantum-to-classical crossover, highlighting the rich and subtle interplay between topology and many-body correlations.

This paper is organized as follows. In the next section, we introduce the Hamiltonians of the two-site and three-site SSH models in the XXZ spin representation, including their interacting extensions. In the Results section, we present our findings on the entanglement entropy across parameter regimes, local magnetization profiles, and the finite-size scaling behavior near phase boundaries. We conclude with a summary of our main results and their implications. Additional technical details, including the implementation of the DMRG calculations and the single-particle band structures, are provided in the Appendix.

\section{Generalized SSH Models with Interactions and Variable
Unit Cell Size}

The Su--Schrieffer--Heeger (SSH) model is a foundational one-dimensional lattice model that illustrates the concept of symmetry-protected topological phases. To study interaction effects, we work with its spin-chain representation,
\begin{align}
    H &= \sum_{n \in A} \left( J_1^{xy}\,(\sigma_n^x \sigma_{n+1}^x + \sigma_n^y \sigma_{n+1}^y ) + J_1^z\, \sigma_n^z \sigma_{n+1}^z \right) \nonumber \\
      &\quad + \sum_{n \in B} \left( J_2^{xy}\,(\sigma_n^x \sigma_{n+1}^x + \sigma_n^y \sigma_{n+1}^y ) + J_2^z\, \sigma_n^z \sigma_{n+1}^z \right) \nonumber \\
      &\quad \vdots \nonumber \\
      &\quad + \sum_{n \in N} \left( J_N^{xy}\,(\sigma_n^x \sigma_{n+1}^x + \sigma_n^y \sigma_{n+1}^y ) + J_N^z\, \sigma_n^z \sigma_{n+1}^z \right),
\end{align}
where the sums over $A,B,\ldots,N$ index the alternating bonds within each unit cell. In the absence of the longitudinal Ising terms $J^z$, this Hamiltonian can be mapped via a Jordan--Wigner transformation to free fermionic SSH model, and is therefore exactly solvable. The transverse couplings $J^{xy}_{1,2,\ldots}$ encode the alternating hopping amplitudes, while the longitudinal couplings $J^z_{1,2,\ldots}$ inherit the same modulation. The sign of $J_z$ determines the nature of the interactions: $J_z>0$ corresponds to repulsive interactions in the fermionic language, mapping to an antiferromagnetic chain, whereas $J_z<0$ corresponds to attractive interactions, mapping to a ferromagnetic chain. In the regime of large $\lvert J_z\rvert$, one expects classical (anti)ferromagnetic order to dominate the ground-state physics.

We now specialize to two generalized SSH models: one with a 2-site unit cell and one with a 3-site unit cell. In both cases, we investigate the entanglement properties of the ground state as a function of the dimerization parameter \( \delta \), both in the presence and in the absence of interactions.

\subsection{2-Site Unit Cell interacting SSH Model}

We consider a parametrization of the SSH spin chain chain given by:
\begin{align}   
    J^{xy}_1 = (1 + \delta), \quad J^{xy}_2 = (1 - \delta),\nonumber \\
    \quad J^z_1 = J_z(1 + \delta), \quad J^z_2 = J_z(1 - \delta).
\end{align}
This gives the spin Hamiltonian in equation 4.
\begin{align}
    H &= \sum_{n \in A} \left( J_1^{xy} \left( \sigma_n^x \sigma_{n+1}^x + \sigma_n^y \sigma_{n+1}^y \right) + J_1^z \sigma_n^z \sigma_{n+1}^z \right) \nonumber \\
        &+ \sum_{n \in B} \left( J_2^{xy} \left( \sigma_n^x \sigma_{n+1}^x + \sigma_n^y \sigma_{n+1}^y \right) + J_2^z \sigma_n^z \sigma_{n+1}^z \right)
\end{align}

In the non-interacting limit (\( J_z = 0 \)), the system reduces to a free-fermion model with two bands. At half filling, the Fermi level lies at \( E = 0 \). When \( \delta > 0 \), the system favors dimerization, gapping out the edge states. When \( \delta < 0 \), two zero-energy edge modes emerge, and the Fermi level coincides with them, making these edge modes the degenerate ground state.

To study the entanglement structure, we compute the bipartite von Neumann entanglement entropy of the ground state. We divide the system into two halves of length \( L = N/2 \), as shown schematically in Fig.~\ref{figure:Model}(a).

\subsection{3-Site Unit Cell interacting SSH Model}

We next consider a three-site unit cell with hopping pattern with parametrization:
\begin{align}
    &J^{xy}_1 = (1 + \delta), \quad J^{xy}_2 = (1 - \delta), \quad J^{xy}_3 = 1, \nonumber \\
    &J^z_1 = J_z(1 + \delta), \quad J^z_2 = J_z(1 - \delta), \quad J^z_3 = J_z.
\end{align}
The corresponding spin Hamiltonian is:
\begin{align}
    H = &\sum_{i=0}^{N/3 - 1} \Big[ J^{xy}_1 \left( \sigma_{3i+1}^x \sigma_{3i+2}^x + \sigma_{3i+1}^y \sigma_{3i+2}^y \right) + J^z_1 \sigma_{3i+1}^z \sigma_{3i+2}^z \nonumber \\
    &+ J^{xy}_2 \left( \sigma_{3i+2}^x \sigma_{3i+3}^x + \sigma_{3i+2}^y \sigma_{3i+3}^y \right) + J^z_2 \sigma_{3i+2}^z \sigma_{3i+3}^z \nonumber \\
    &+ J^{xy}_3 \left( \sigma_{3i+3}^x \sigma_{3i+4}^x + \sigma_{3i+3}^y \sigma_{3i+4}^y \right) + J^z_3 \sigma_{3i+3}^z \sigma_{3i+4}^z \Big].
\end{align}

In the absence of interactions, this model yields three bands. Unlike the two-site model, the three-site model lacks inversion symmetry, and the edge state is not symmetry-protected. A single edge mode forms at the interface between the lower and middle bands. For \( \delta > 0 \), this edge mode is localized at the left end; for \( \delta < 0 \), at the right end. Across all values of \( \delta \), the edge state always appears near the \( 2N/3 \)-th energy level. Thus, the edge state becomes the ground state at \emph{two-thirds filling}, where \( 2/3 \) of the spins are up and \( 1/3 \) are down. For a discussion of the full phase diagram of the general 3-site unit cell SSH model, please refer to Ref. \cite{PhysRevB.101.235155}.

\begin{figure}
\subcaptionbox{\protect\label{a}}{\includegraphics[width=0.5\textwidth]{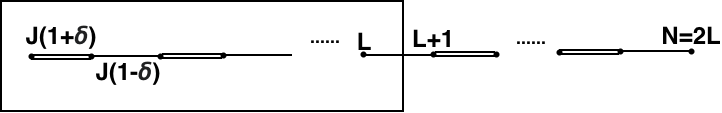}}\hfill
\subcaptionbox{\protect\label{b}}
{\includegraphics[width=0.5\textwidth]{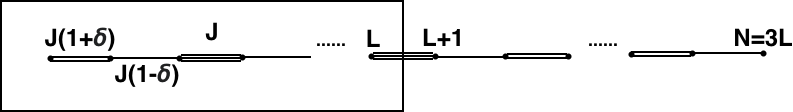}}\hfill
\caption{
     Illustration of the generalized SSH chains with (a) a 2-site unit cell and (b) a 3-site unit cell. The bonds represent cyclically alternating couplings: \( J_1, J_2 \) in (a), and \( J_1, J_2, J_3 \) in (b). The boxes indicate the bipartition of the system used to compute the von Neumann entanglement entropy of the ground state. In (a), the system is cut at the midpoint (\( L = N/2 \)), while in (b), the cut is placed on a \( J_3 \) bond at subsystem length \( L = N/3 \).
}
\label{figure:Model}
\end{figure}

To study entanglement, we define the subsystem length \( L = N/3 \) and place the bipartition cut on the \( J^{xy}_3 = 1 \) bond, preserving the unit cell structure (see Fig.~\ref{figure:Model}(b)).

\section{DMRG Analysis of Generalized SSH Chains with Interactions}

\subsection{The 2-site unit cell interacting SSH model}

To investigate the effects of topology in the interacting SSH model with a two-site unit cell, we performed density matrix renormalization group (DMRG) calculations using the \texttt{tenpy} package. The total spin number was conserved to maintain half-filling, ensuring that edge modes in topologically non-trivial regimes appear in the ground state and are thus accessible via DMRG.

\subsubsection{Entanglement Entropy}

We explored the parameter space by sweeping over the dimerization $\delta$ and interaction strength $J_z$, and measured the entanglement entropy (EE) at each point. At critical points, where the system undergoes a phase transition, we observe strong finite-size scaling behavior, whereby the EE diverges with system size.

\begin{figure}[h]
\centering
\includegraphics[width=1.0\linewidth]{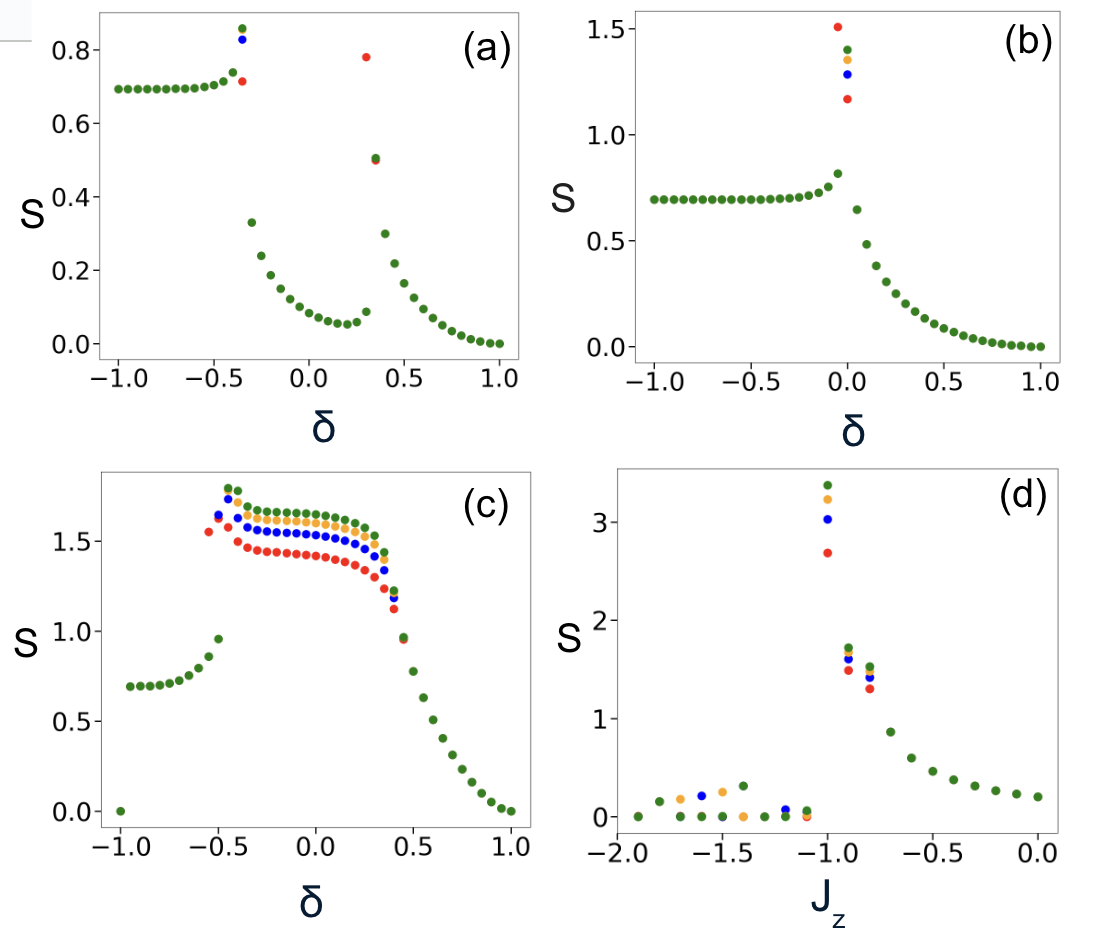}
\caption{Bipartite von Neumann entanglement entropy in the interacting 2-site unit cell SSH model evaluated along various cuts in parameter space spanned by $( J_z, \delta)$. Here we consider $\delta$ sweeps at fixed interaction strengths, where each color represents one fixed strength(Red: L=100, Blue: L=200, Orange: L=300, Green: L=400)(a) $J_z$=4.0 (strong repulsive interaction) (b) $J_z$=0.0 (no interaction) (c) $J_z$=-0.8 (moderate attractive interaction. In  (d) we fix $\delta$=0.30 and sweep through $J_z$. The observed singularities allow us to map out the phase diagram shown in Fig. ~\ref{figure:Phase Diagram}.}
\label{figure:sweep_SSH}
\end{figure}

Our results indicate that for sufficiently strong repulsive interactions $J_z > 1.0$, two distinct transitions emerge, as shown in Fig.~\ref{figure:sweep_SSH}(a) for $J_z = 4.0$. In conjunction with the magnetization profiles discussed in the next subsection, this implies the presence of an intermediate renormalized classical antiferromagnetic phase between the two non-magnetic topologically distinct phases that are already known from the non-interacting case. Notably, $J_z = 1.0$ corresponds to the SU(2)-symmetric point.

When $J_z$ lies in the range $-\frac{1}{\sqrt{2}} < J_z < 1$, only a single transition is observed at $\delta = 0$, consistent with the non-interacting case. The observation that there is such a finite regime in parameter space centered around ($J_z=0$) is a manifestation of topological protection that is only overcome when the interaction $J_z$ exceeds a critical strength whose magnitude is set by the topological gap of the non-interacting system.

As $J_z$ is decreased further into the range $-1 < J_z < -\frac{1}{\sqrt{2}}$, the single transition point broadens into a finite region exhibiting non-trivial scaling behavior, as seen in Fig.~\ref{figure:sweep_SSH}(c). This will be discussed in more detail below. A similar analytical result was found in \cite{PhysRevB.24.5229}, where the free energy stops behaving singularly. Moreover, in the renormalization group language, the flow inside the critical regime is in a different direction from the flow outside of the regime.

For $J_z < -1$, the entanglement entropy vanishes, indicating a gapped, classical ferromagnetic phase. This behavior is clearly visible in Fig.~\ref{figure:sweep_SSH}(d), where we sweep over the negative $J_z$ region at fixed $\delta = 0.30$. The plot reveals three distinct regimes: a ferromagnetic phase with vanishing EE for $J_z < -1$, an intermediate region with significant finite-size scaling of the EE, and a third phase where this scaling disappears as $J_z$ increases.

\subsubsection{Magnetization Profiles}

As discussed in the previous section, three distinct regimes emerge when $J_z > 1$, and two additional regimes appear when $J_z < -\frac{1}{\sqrt{2}}$. One of these exhibits finite-size scaling behavior, whereas the other is classical and therefore has  vanishing entanglement entropy (EE). To better understand the quantum behavior characterizing each regime, we analyze the site-dependent expectation value of the magnetization, denoted as $\langle S_z \rangle$.

\begin{figure}
\subcaptionbox{\protect\label{a}}{\includegraphics[width=0.48\linewidth,height=0.42\linewidth]{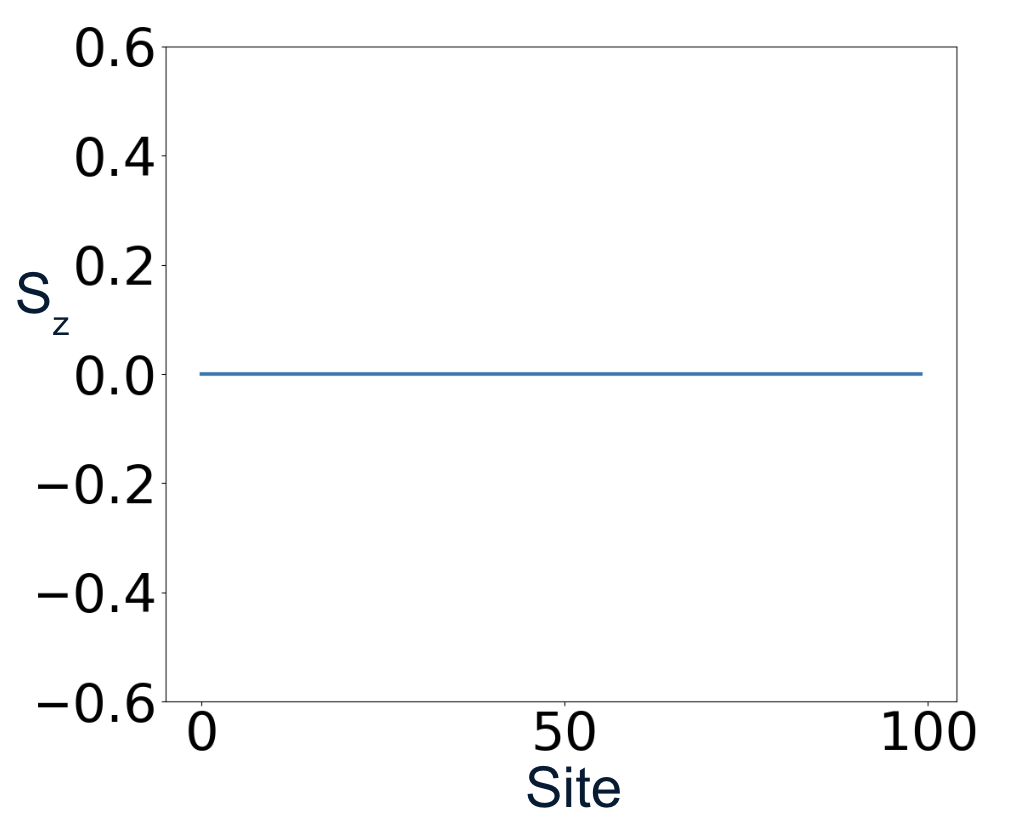}}\hfill
\subcaptionbox{\protect\label{b}}
{\includegraphics[width=0.48\linewidth,height=0.42\linewidth]{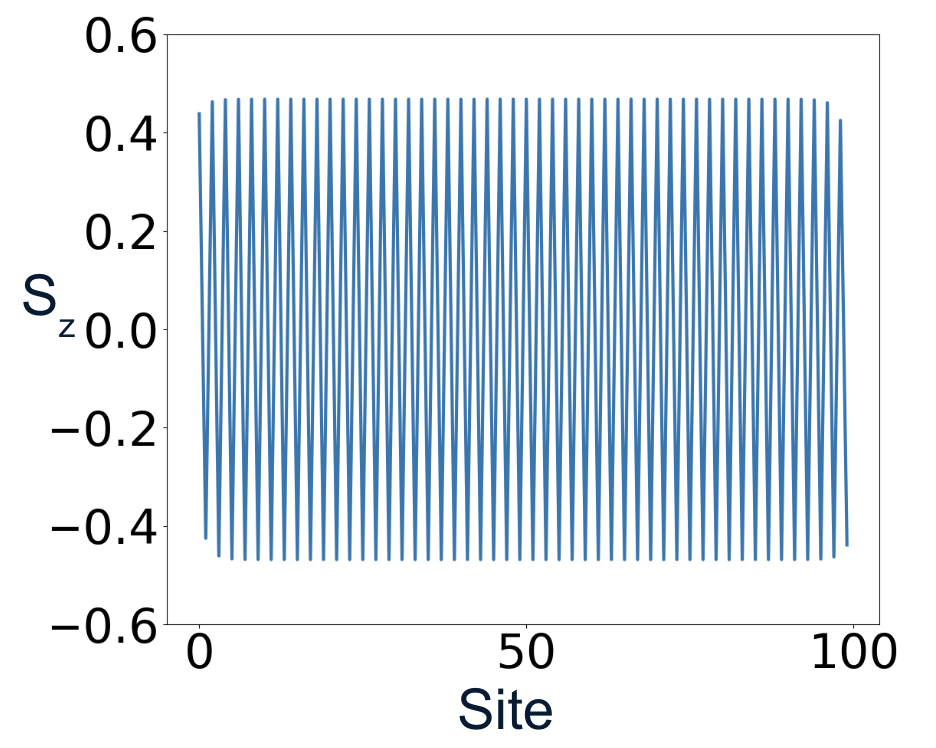}}\hfill

\centering
\subcaptionbox{\protect\label{c}}
{\includegraphics[width=0.48\linewidth,height=0.42\linewidth]{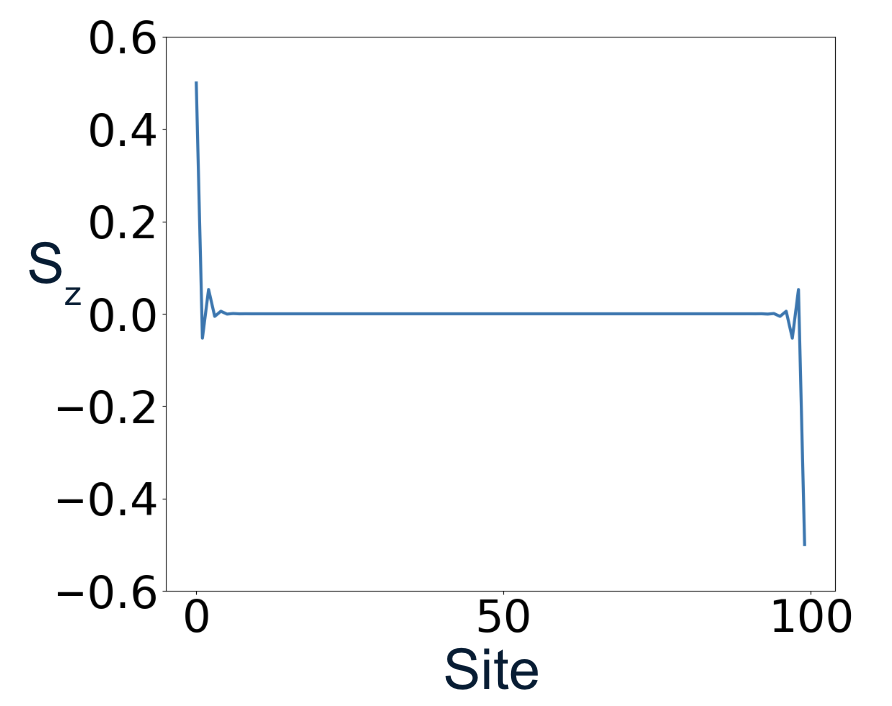}}\hfill
\caption{Magnetization profiles in the two-site unit cell SSH model with repulsive interactions. Here we show the local  expectation values of the magnetization $\sigma^z_i$ at each site for (a) $J_z$=4.0, $\delta$=-0.9 (topologically non-trivial regime with surface states) (b)$J_z$=4.0, $\delta$=0.0 (classical antiferromagnetic regime) (c)$J_z$=4.0, $\delta$=0.9 (topologically trivial regime).}
\label{Magnetization_SSH}
\end{figure}

Fig. \ref{Magnetization_SSH} displays $\langle \sigma_z \rangle$ for the three phases observed at $J_z = 4.0$. When $\delta$ is strongly negative, the system behaves similarly to the $\delta < 0$ regime at $J_z < 1$: the two end sites decouple from the bulk, forming localized surface states, whereas the remaining sites pair into singlets, resulting in $\langle \sigma_z \rangle \approx 0$ throughout the bulk.
For strongly positive $\delta$, the intra-cell coupling dominates, leading to uniform dimerization across the chain. This again results in vanishing magnetization on all sites.

In the intermediate phase, the chain exhibits antiferromagnetic ordering, with alternating up and down spins along the chain. This behavior can be understood from the dominance of the $\sigma_z \otimes \sigma_z$ interaction: for positive $J_z$, energy is minimized when neighboring spins are anti-aligned. Close to the phase boundaries this sublattice magnetization is observed to be reduced by fluctuations. In the opposite limit, i.e., as the interaction term becomes dominant, the system becomes increasingly classical, and therefore the entanglement vanishes. 

For $J_z<-\frac{\sqrt{2}}{2}$ in the regime with critical behavior with non-universal finite-size scaling behavior, the chain remains nearly dimerized, and the magnetization remains close to zero at all sites.
When $J_z <-1$, the system favors fully classical ferromagnetic alignment due to the attractive interaction. In this regime, parallel spin alignment lowers the energy, resulting in a finite, uniform magnetization across the chain.

\subsubsection{Finite-Size Scaling and Central Charge}

In addition to characterizing the distinct regimes, we have also investigated the nature of the phase transitions between them via finite size scaling analysis.  For a quantum spin chain with open boundary conditions, the entanglement entropy $S$ as a function of subsystem size $L$ at criticality follows the scaling relation \cite{Calabrese2004}:
\begin{equation}
    S = \frac{c}{6} \ln\left(\frac{N}{\pi} \sin\left(\frac{\pi L}{N}\right)\right) + \Delta S(L),
\end{equation}
where $c$ is the central charge and $N$ is the total system size. The higher order correction term $\Delta S(L)$ is only relevant in the region with critical behavior, as discussed below. By fitting numerical DMRG data for the entanglement entropy $S$ versus $\ln L$, we can extract the central charge of the corresponding conformal field theory and thus identify the universality classes of the various transitions.

\begin{figure}
\subcaptionbox{\protect\label{a}}{\includegraphics[width=0.48\linewidth,height=0.42\linewidth]{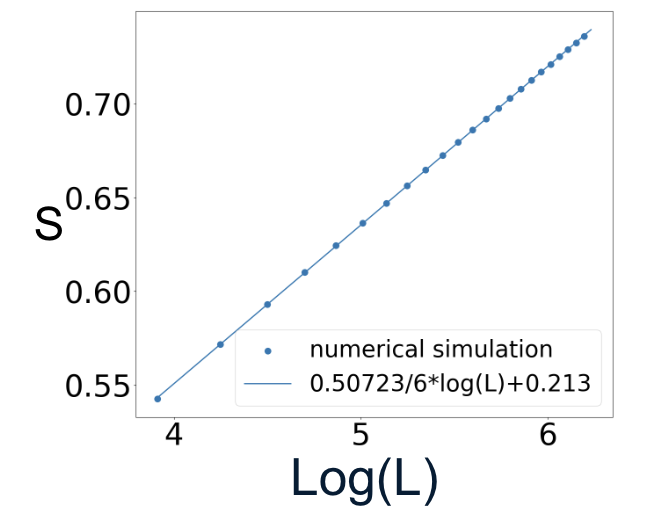}}\hfill
\subcaptionbox{\protect\label{b}}
{\includegraphics[width=0.48\linewidth,height=0.42\linewidth]{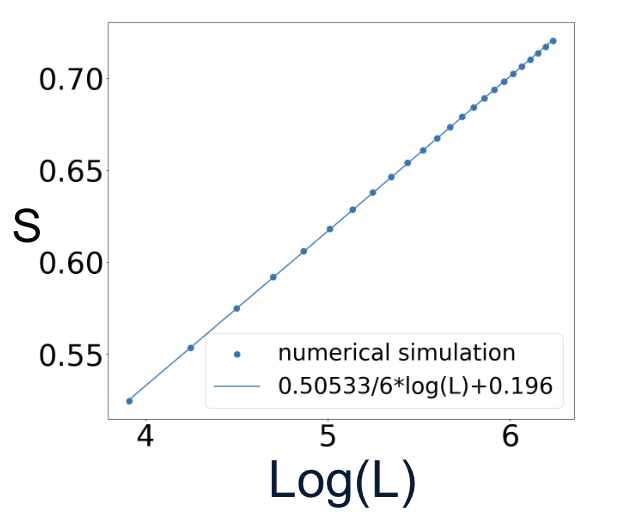}}\hfill
\subcaptionbox{\protect\label{c}}
{\includegraphics[width=0.48\linewidth,height=0.42\linewidth]{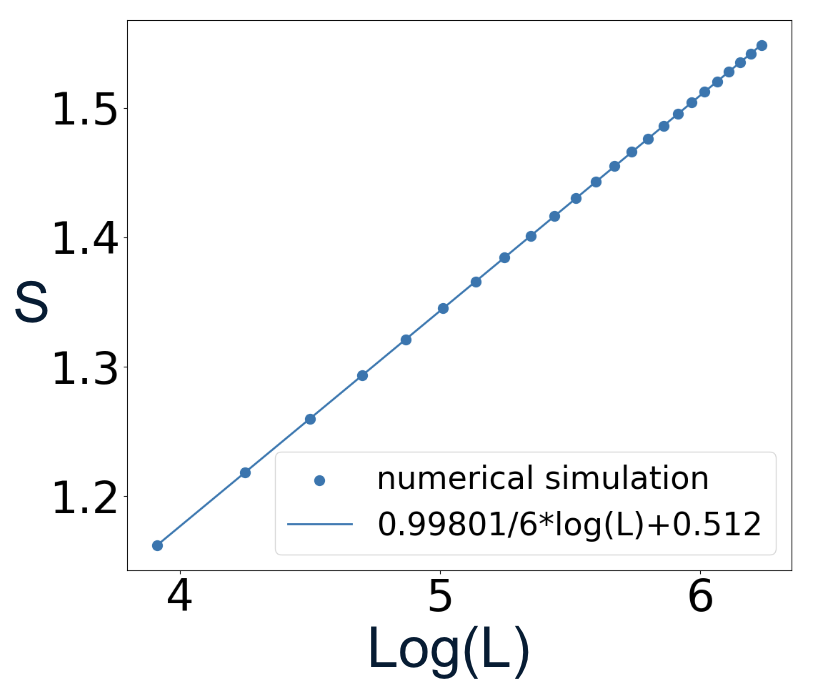}}\hfill
\subcaptionbox{\protect\label{d}}
{\includegraphics[width=0.48\linewidth,height=0.42\linewidth]{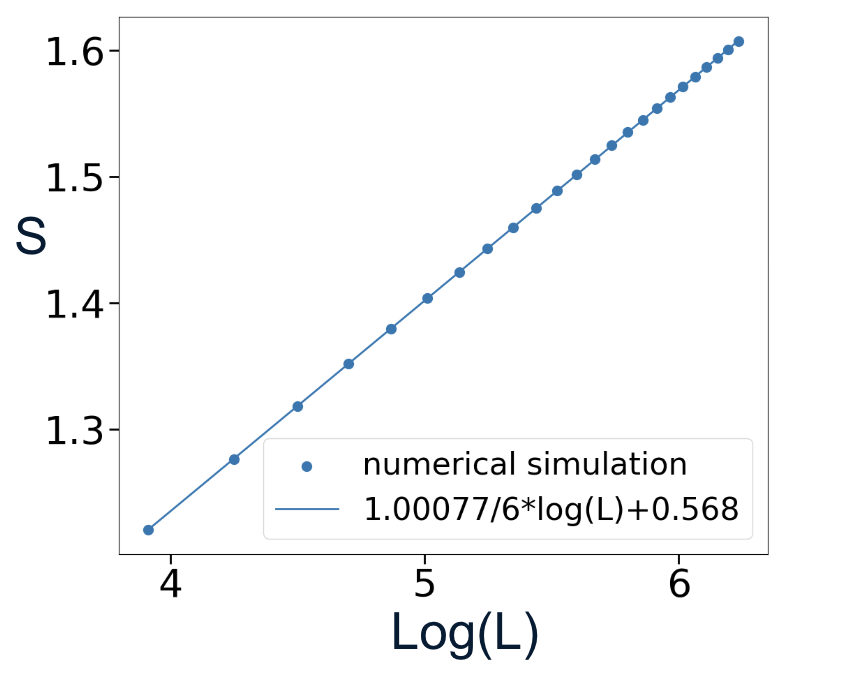}}\hfill
\caption{The scaling of S with L at critical point for (a) $J_z$=4.0, $\delta$=0.3415 (b) $J_z$=-0.5, $\delta$=0.0 (c) $J_z$=4.0, $\delta$=-0.3415, and  (d) $J_z$=-0.74, $\delta$=0.10.}
\label{Scaling S SSH}
\end{figure}

This analysis reveals the following central charges at the critical lines:
\begin{itemize}
\item For the two transitions at $J_z > 1$, the extracted central charge is $c = 0.5$.
\item For the transition occurring at intermediate $J_z$ between $-\frac{1}{\sqrt{2}}$ and 1, $c = 1$.
\item Throughout the Kosterlitz-Thouless transition line and the regime of critical behavior in $-1<J_z<-1/\sqrt{2}$ , we also find $c = 1$.
\end{itemize}
These results are shown in the entanglement scaling plots in Figs.~\ref{Scaling S SSH}(a),(b),(c),(d). A more detailed analytical classification of these transitions, including associated central charge values, can be found in \cite{PhysRevB.24.5229}.

We now discuss the physical interpretation of each transition:
\begin{itemize}
    \item When $J_z > 1$, the system exhibits an intermediate classical antiferromagnetic phase, bounded by two critical lines beyond which the system transitions to a dimerized state\cite{PhysRevB.54.9862}. These transitions are  consistent with Ising criticality, which is characterized by a central charge $c = \frac{1}{2}$ \cite{PhysRevLett.96.100603, PhysRevLett.90.227902}.
    \item Along the blue line at $\delta = 0$ and $-\frac{1}{\sqrt{2}} < J_z < 1.0$, the critical behavior is described by a gapless Luttinger liquid with c=1. As $J_z \rightarrow 1$, the system becomes SU(2) symmetric, described by a k=1 Wess-Zumino-Witten conformal field theory with a conformal charge $c = 3k/(k+2)=1$ \cite{PhysRevLett.47.1840, PhysRevB.34.6372, PhysRevB.102.180409}.
    \item The regime with critical behavior below the green transition line  in the phase diagram is characterized by a diverging correlation length $\xi\sim\exp(A/\sqrt{|K-K_c|})$. This  Kosterlitz-Thouless (KT) transition arises due to the negative $J_z$ interaction that twists the spin orientation. Using bosonization, valid for sufficiently small $\delta$ and $J_z$, we find $K(J_z,\delta)\approx (1+(J_z/2\pi))(1-a\delta^2)$. This exponent also controls the numerically observed non-universal finite-size corrections to the EE via $\Delta S\sim \cos(\pi(L+1))/L^{2K(J_z,\delta)}$.
    \item At $J_z = -1$, the system enters a gapped phase with broken $\mathbb{Z}_2$ symmetry. This transition from the regime with critical behavior to long-range ferromagnetism is Ising like, and therefore described by a c=1/2 CFT. 
\end{itemize}

Combining the observed transitions, their critical properties, and the quantum behavior in each regime, we construct the full phase diagram shown in Fig.~\ref{figure:Phase Diagram}(a).

\subsection{The 3-site unit cell interacting SSH model}

Let us begin by reviewing the phase diagram of the non-interacting 3-site unit cell SSH model, previously analyzed using an entanglement difference measure~\cite{PhysRevB.101.235155}. As shown in Fig.~\ref{figure:Phase Diagram} for representative coupling configurations, the model exhibits three distinct bulk bands—bonding, non-bonding, and anti-bonding—resulting from the enlarged unit cell. Depending on the hopping amplitudes and boundary conditions, edge states can appear within the bulk gaps between these bands.

At half filling—corresponding to zero magnetization in the spin language—the system remains in a gapless Luttinger liquid phase for all parameter choices. However, the numbers and energies of edge states vary, depending on the intra-unit-cell hopping parameters \( t_1, t_2, t_3 \) and on the boundary terminations. Assuming an open chain whose length is a multiple of three and with hopping terms arranged cyclically as \( t_1 \text{--} t_2 \text{--} t_3 \), the number of edge-state pairs behaves as follows (see also Fig.~2 and Table~1 in~\cite{PhysRevB.101.235155}):

\begin{itemize}
    \item Zero edge states when \( t_1, t_2 > t_3 \).
    \item One edge-state pair in two distinct regions:
         \( t_2 > t_1 > t_3 \) and 
        \( t_1 > t_2 > t_3 \). These are either localized on the left or the right end of the open chain, depending on whether the adjacent bond ($t_1$ or $t_2$) is the weakest.
    \item Two edge-state pairs when \( t_1, t_2 < t_3 \), localized on both ends of the open chain. 
    \item No edge states and bulk gap closing along the singular line \( t_1 = t_2 = t_3 \), corresponding to a uniform chain.
\end{itemize}

In the parametrization used here, $t_1=(1+\delta), t_2=(1-\delta), t_3=1$ , scanning through the modulation parameter \( \delta \) takes the system through the two sectors with one edge-state pair, separated by the gapless point at \( \delta = 0 \) with no edge states.

In analogy with the standard anisotropic Heisenberg chain, the addition of an Ising interaction term in the spin language is expected to drive the system through Kosterlitz-Thouless  transitions at half filling. These occur at the isotropy points \( J_z = \pm J_{xy} \), beyond which the system becomes gapped and enters either a ferromagnetic or antiferromagnetic phase. 

However, a key difference from the standard SSH model lies in the richer high energy structure of the three-band system. In particular, the presence of multiple gaps and the distribution of in-gap edge states offer an opportunity to explore how interactions affect  localized modes at energies $E\ne 0$. In the following sections, we analyze these edge states in the presence of repulsive and attractive interactions.

In DMRG, to  study the effects of interactions on the edge modes in the three-site SSH chain, we consider the system at two-third filling, which in the spin language means that we are in a $S^z_{tot}$ sector where two-thirds of the spins are up and one-third are down. In this case, the in-gap edge states located between the non-bonding and the anti-bonding band become the ground states of the system. 

\subsubsection{Entanglement Entropy}

\begin{figure}[H]
\centering
\includegraphics[width=1.0\linewidth]{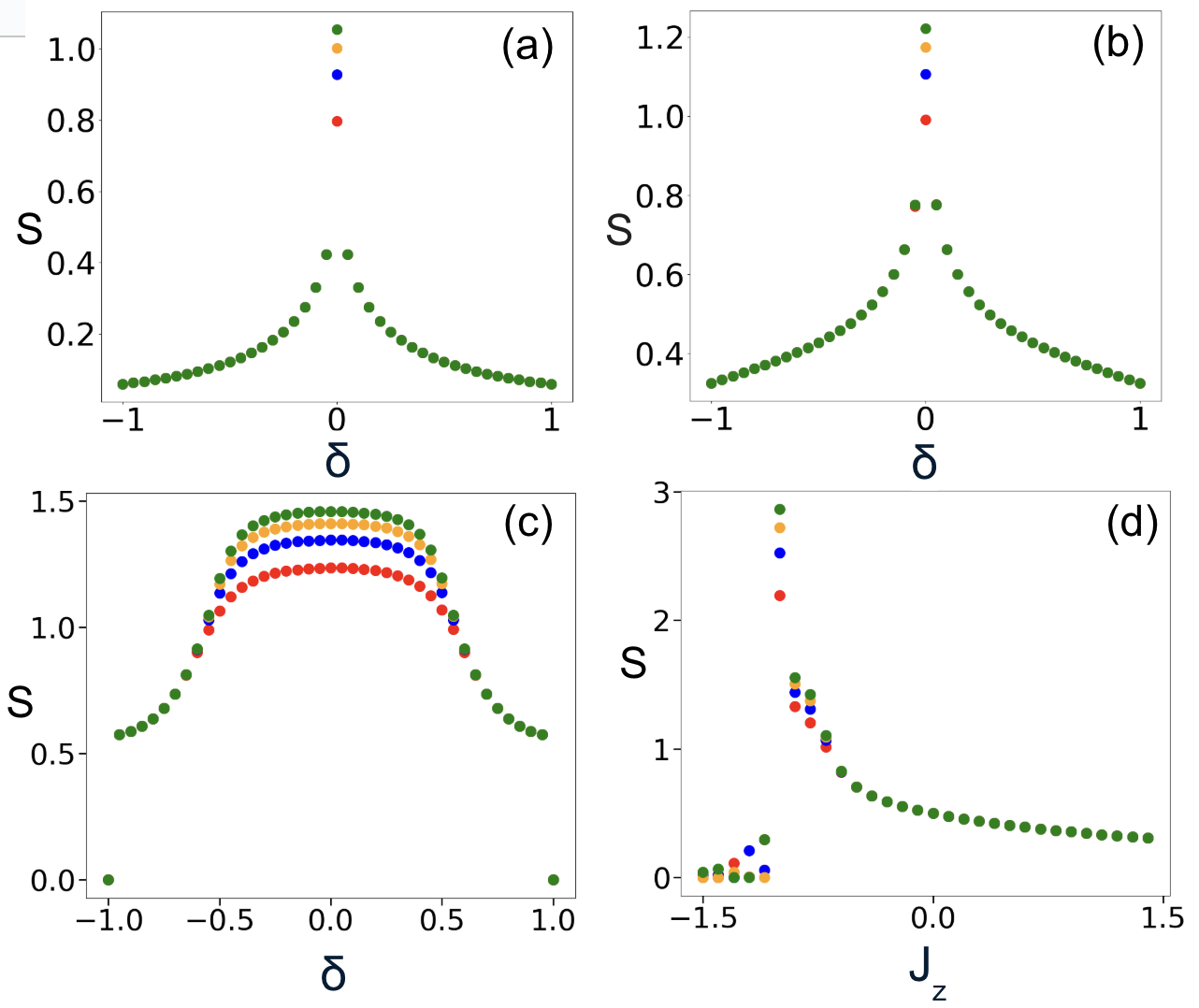}
\caption{Bipartite von Neumann entanglement entropy in the interacting 3-site unit cell SSH model at 2/3 filling, evaluated along various cuts in parameter space spanned by $( J_z, \delta)$. Here we consider $\delta$ sweeps at fixed interaction   strengths (a) $J_z$=4.0  (b) $J_z$=0.0  (c) $J_z$=-0.8. In  (d) we fix $\delta$=-0.30 and sweep through $J_z$. The observed singularities allow us to map out the phase diagram shown in Fig. 8. Each color represents one fixed strength(Red: L=30, Blue: L=60, Orange: L=90, Green: L=120).
}
\label{Sweep SSH3}
\end{figure}

Analogous to the previous discussion of the two-site SSH case, we perform a parameter sweep over the interaction strength \( J_z \) and the dimerization parameter \( \delta \), and measure the bipartite von Neumann entanglement entropy at each point, using the bipartition cut shown in Fig. \ref{figure:Model}(b). The results are shown in Fig.~\ref{Sweep SSH3}.

In contrast to the two-site unit cell system, the entire region \( J_z > -\frac{1}{\sqrt{2}} \) exhibits only a single transition at \( \delta = 0 \), separating two low-entropy (classical) phases where the edge states switch from localization on the left boundary to the right boundary. No additional intervening phase is found for large positive \( J_z \). This observation is consistent with the fact that the addition of Ising interactions at this filling is a relevant perturbation, but the antiferromagnetic order favored by the Ising interaction cannot be accommodated in the $\sigma^z_{tot}=2/3$ sector.

However, similar to the two-site case, in the regime of negative \( J_z \), we  find a finite region of critical behavior with strong finite-size corrections to the EE for \( -1 < J_z < -\frac{1}{\sqrt{2}} \), and a gapped regime with vanishing EE for \( J_z < -1 \), consistent with ferromagnetic ordering.

\subsubsection{Magnetization Profiles}

\begin{figure}[H]
\subcaptionbox{\protect\label{a}}{\includegraphics[width=0.48\linewidth,height=0.42\linewidth]{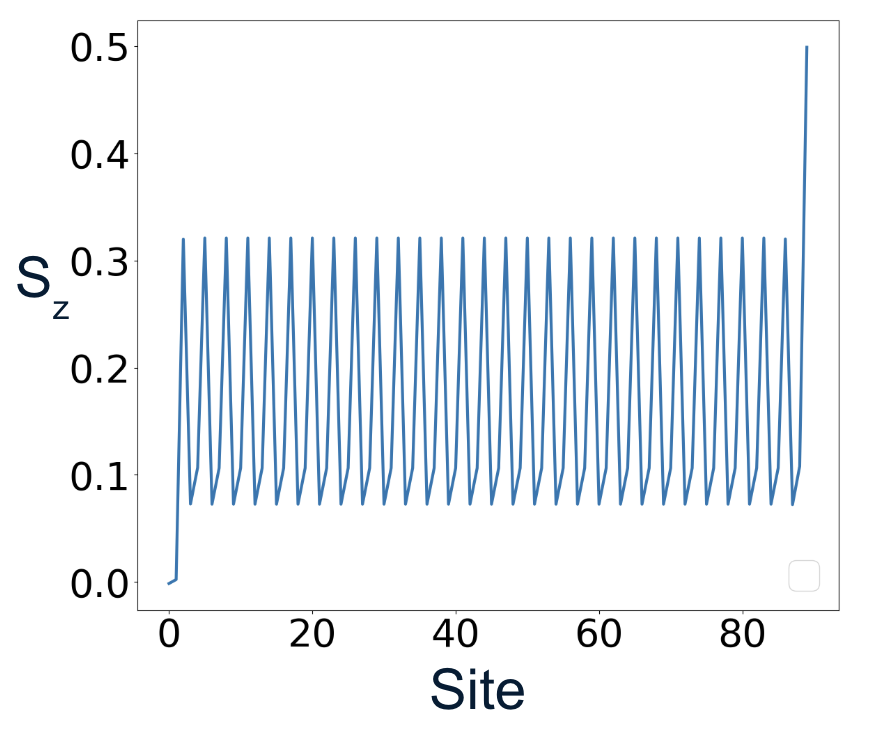}}\hfill
\subcaptionbox{\protect\label{b}}
{\includegraphics[width=0.48\linewidth,height=0.42\linewidth]{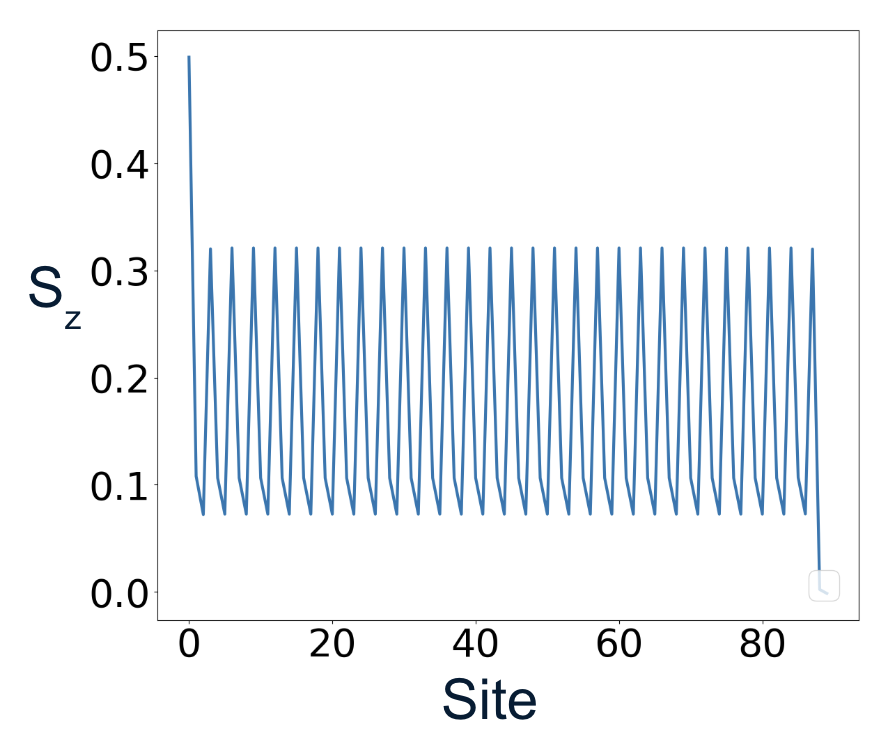}}\hfill
\caption{Magnetization profiles in the three-site unit cell SSH model at 2/3 filling with slightly attractive interactions. Here we show the local  expectation values of the magnetization $\sigma^z_i$ at each site for (a) $J_z$=-0.5, $\delta$=-0.95, and (b) $J_z$=-0.5, $\delta$=0.95.}
\label{Magnetization SSH3}
\end{figure}

We next examine the expectation value of the local magnetization \( \langle S_z \rangle \) at each site to better characterize the different quantum phases. The results are presented in Fig.~\ref{Magnetization SSH3}. 

Because of the non-zero filling, corresponding to a finite applied magnetic field along the z-direction in the spin language, there is a non-vanishing local magnetization on all lattice sites. In addition, due to the antiferromagnetic Ising interactions, there is a finite staggered magnetization. Nonetheless, there are still interesting edge phenomena observable in this regime. 

Specifically, for all \( J_z > -\frac{1}{\sqrt{2}} \), edge 
localization is observed in the magnetization profiles on one of the two open chain ends, consistent with the non-interacting case.  When \( \delta > 0 \), the rightmost site has the highest magnetization, indicating an edge mode localized on the right. Conversely, for \( \delta < 0 \), the edge mode shifts to the leftmost site. In addition, unlike the half-filled case, where bulk sites tend to exhibit zero magnetization, the two-thirds filling leads to finite bulk magnetization due to the overall spin imbalance.

Along the critical line \( -1 < J_z < -\frac{1}{\sqrt{2}} \), the magnetization is uniform throughout the chain, with an average value around 0.16. This reflects the shift in the effective chemical potential caused by the spin imbalance. For \( J_z < -1 \), the system enters a ferromagnetic phase dominated by the \( S_z \)-\( S_z \) coupling, leading to uniform alignment of spins.

\subsubsection{Finite-Size Scaling and Central Charge}

As previously discussed in the context of the two-site unit cell system, at criticality, the EE is expected to follow a logarithmic scaling with system size, described by  conformal field theory via Eq. 7.
Again, we fit DMRG data for the EE as a function of \( \ln L \) at representative critical points, as shown in Fig.~7(a)-(c). However, in this case the bipartition cut is taken at one third of the chain length (Fig. \ref{figure:Model}(b)).

\begin{figure}
\subcaptionbox{\protect\label{a}}{\includegraphics[width=0.48\linewidth,height=0.42\linewidth]{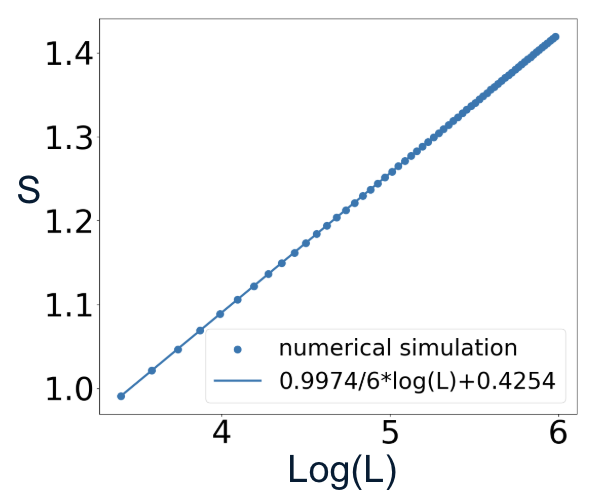}}\hfill
\subcaptionbox{\protect\label{b}}
{\includegraphics[width=0.48\linewidth,height=0.42\linewidth]{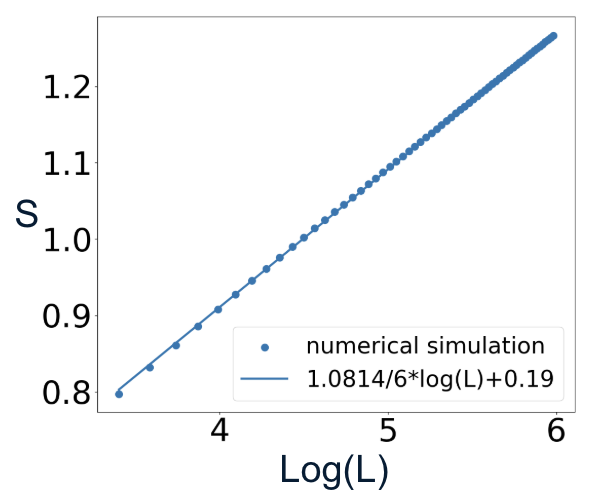}}\hfill
\centering
\subcaptionbox{\protect\label{c}}
{\includegraphics[width=0.48\linewidth,height=0.42\linewidth]{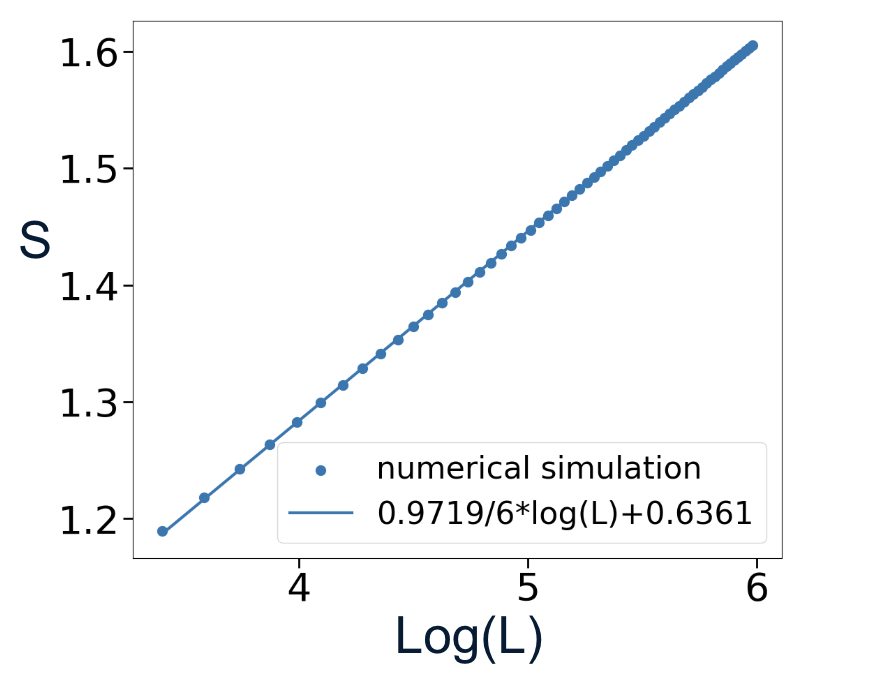}}\hfill
\caption{The fitting of S with respect to L for (a)$J_z$=0.0, $\delta$=0.0 (b)$J_z$=4.0, $\delta$=0.0 (c)$J_z$=-0.74, $\delta$=0.1}
\label{Scaling S SSH3}
\end{figure}

Figs. \ref{Scaling S SSH3}(a) and (b) show the scaling at the transition between left- and right-localized edge modes for \( J_z > -\frac{1}{\sqrt{2}} \). At \( J_z = 1 \), the central charge is approximately \( c = 1 \), consistent with Luttinger Liquid behavior. However, for larger \( J_z \) (e.g., \( J_z = 4 \)), the numerically extracted central charge increases above 1.

At the phase boundary between the region of critical behavior and the gapped regime (\( J_z > -\frac{1}{\sqrt{2}}\)), the central charge is 1, analogous to  the Kosterlitz Thouless transition observed in  the two-site SSH model [see Fig.~\ref{Scaling S SSH3}(c)].

\begin{figure*}
    \subcaptionbox{\protect\label{a}}{\includegraphics[width=0.48\textwidth,height=0.42\textwidth]{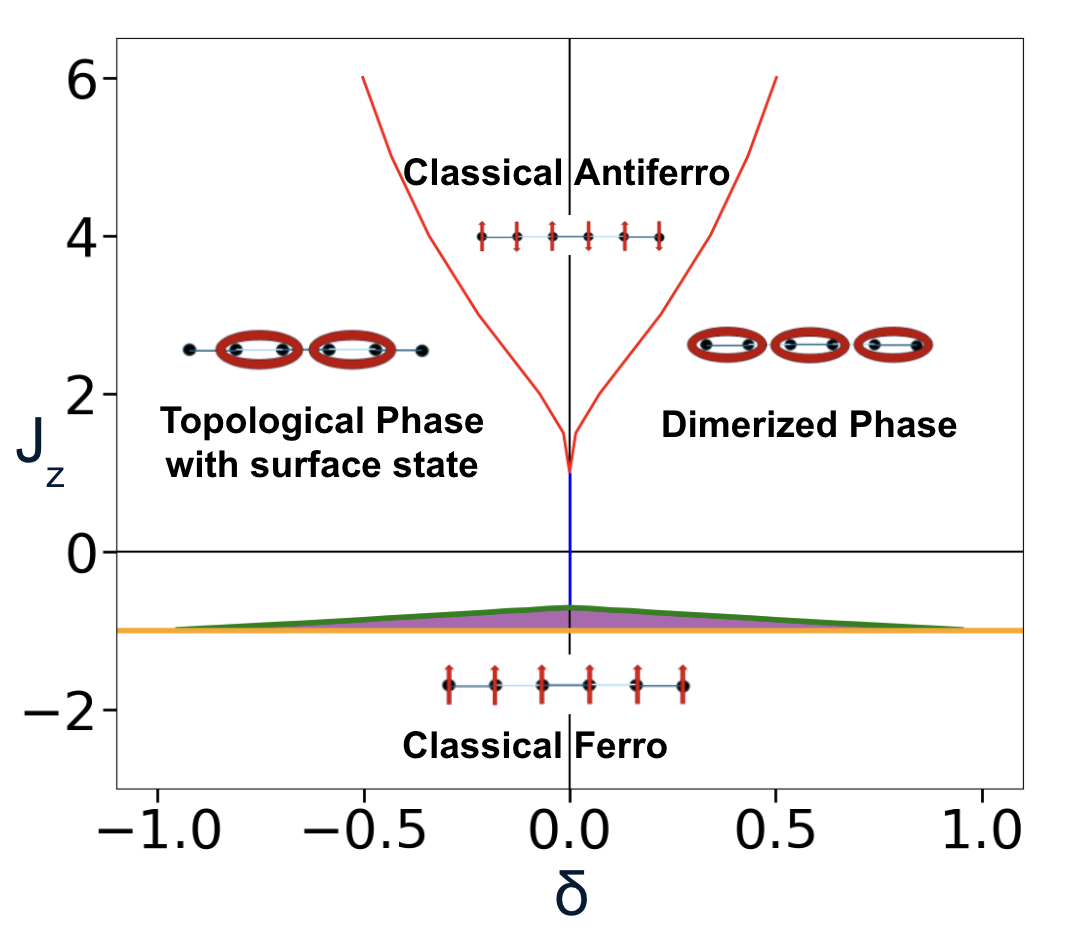}}\hfill
\subcaptionbox{\protect\label{b}}
{\includegraphics[width=0.48\textwidth,height=0.42\textwidth]{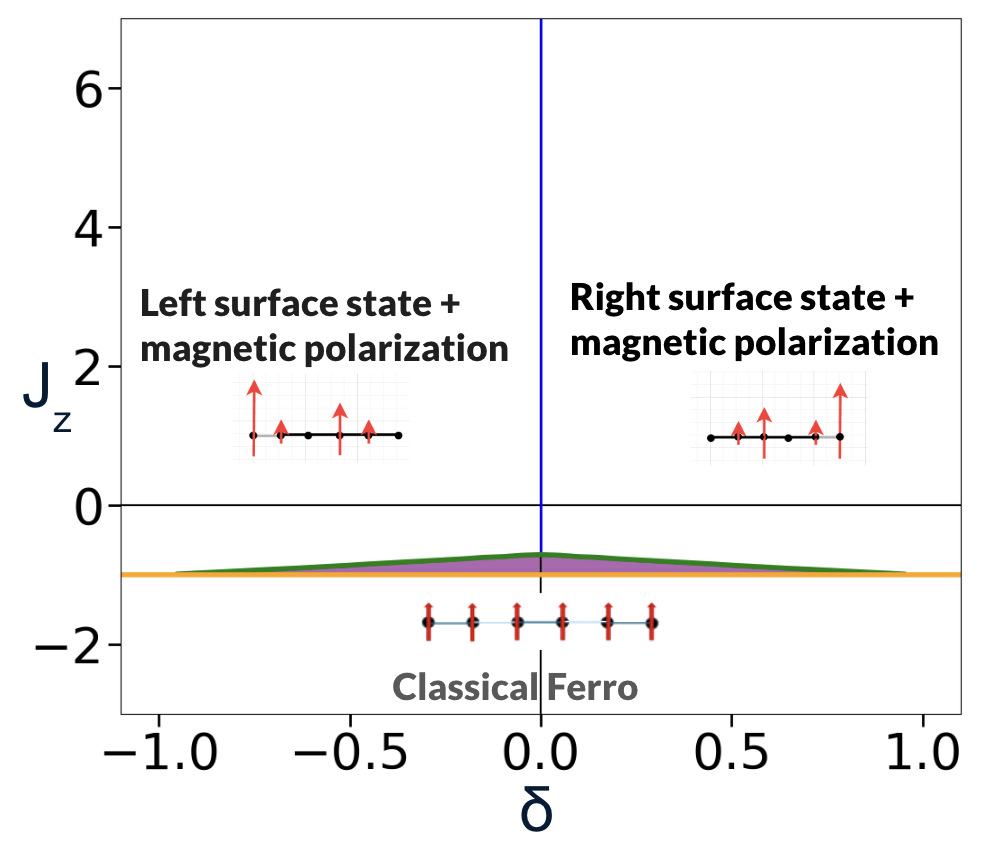}}\hfill
\caption{Phase diagrams of (a) the 2-site unit cell SSH model (b) the 3-site unit cell SSH model. Each line represents different types of transition, where red is Ising Criticality, Blue is the line with Luttinger Liquid Behavior, Green represents Kosterlitz-Thouless transition, Orange represents Anistropic Driven Ferromagnetic Transition, and the shaded region is the region with critical behavior}
\label{figure:Phase Diagram}
\end{figure*}

Combining these observations, we construct the phase diagram shown in Fig.~\ref{figure:Phase Diagram}(b). Notably, in this model, no intervening  phase appears at large positive \( J_z \). Instead, we observe an extended Luttinger liquid regime extending to arbitrarily large positive \( J_z \). For negative \( J_z \), the phase boundaries and behaviors are qualitatively similar to those in the two-site model.

\section{Conclusions}

In this study, we have investigated the properties of the two-site unit cell and three-site unit cell SSH models using DMRG simulations. To accurately capture their edge mode physics, we selected appropriate filling fractions. By sweeping through the parameter space spanned by the interaction strength \( J_z \) and the dimerization \( \delta \), and by measuring the entanglement entropy and magnetization profile at each point, we were able to identify distinct quantum phases and locate as well as characterize the associated transition lines via finite size scaling analysis. 

For the two-site SSH model at half-filling, the phase diagram contains several regimes. At large positive \( J_z \), the \(\sigma_z\) interaction favors classical antiferromagnetic alignment, giving rise to an intermediate phase---absent in the non-interacting limit---between the topologically trivial and non-trivial regimes. Below \( J_z < -1/\sqrt{2} \), a critical region emerges, characterized by finite-size corrections of the EE and vanishing on-site magnetization, indicative of dimerization. As \( J_z \) is further decreased past \( -1 \), the interaction favors ferromagnetic ordering, and we observe a fully polarized phase.

In the three-site unit cell SSH model, at two-thirds filling, a different phenomenology arises. Unlike the two-site case, there is no intermediate antiferromagnetic phase. The only critical line appears at \( \delta = 0 \). On either side of this line, all sites---including those in the bulk---exhibit nonzero magnetization, consistent with antiferromagnetic order rather than dimerization. In these regimes, we observe localization on the right or left edge of the finite chain. As \( J_z \) increases, these edge modes become less localized, and the system approaches uniform antiferromagnetic behavior. In the negative \( J_z \) regime, the behavior mirrors that of the two-site model: a finite-size scaling region emerges for \( J_z < -1/\sqrt{2} \), followed by a ferromagnetic phase for \( J_z < -1 \). However, the on-site magnetization in this case remains finite due to constraints imposed by the conserved filling in the DMRG calculations.

To classify the nature of the phase transitions, we extracted the central charge from the scaling of EE with subsystem size \( L \). For the two-site model, the critical lines correspond to known values of central charge and are consistent with analytical predictions \cite{PhysRevB.24.5229}. In the three-site model, we find that the central charge along the \( \delta = 0 \) line increases with \( J_z \), while the critical line separating the finite-size scaling and non-scaling regimes exhibits the same universal behavior as in the two-site model.

Interestingly, the response of edge modes to an antiferromagnetic Ising perturbation differs markedly between the two models. In the two-site SSH model, the topological zero-energy edge states are robust against weak interactions and remain protected up to a finite critical \( J_z \). In contrast, in the three-site model, where the edge modes appear at finite energy, this protection is absent. Nonetheless, some degree of localization persists: the perturbed edge states become polarized and retain localization features inherited from the non-interacting case. This suggests that even in the absence of strict topological protection, localized boundary signatures may survive due to the underlying non-interacting structure.

These conclusions can be extended to generalized interacting SSH models. Specifically, systems with an even number of sites per unit cell are expected to exhibit similar behavior to the two-site unit cell chain, including the emergence of intermediate antiferromagnetic phases. In contrast, models with an odd number of sites per unit cell behave analogously to the three-site model studied here, and typically host edge modes with asymmetric localization---i.e., localized only on one end of the chain.

\begin{acknowledgments}
We gratefully acknowledge Hubert Saleur and Henning Schlömer for  valuable  discussions. We also thank the Center for Advanced Research Computing (CARC) at the University of Southern California for providing the computational resources that supported the research presented in this publication. URL: https://carc.usc.edu. SH acknowledges support by the Department of Energy (DOE-ASCR) Award DE-SC0026337.
\end{acknowledgments}

\nocite{*}

\bibliography{apssamp}

\clearpage

\appendix

\section{Numerical Details}
We performed our numerical simulations using the TeNPy package, specifically employing the XXZChain model to construct the system. To conserve total spin, we initialized the MPS in different configurations depending on the filling. At half-filling, we used an alternating spin-up/spin-down pattern, while for two-thirds filling, we initialized the state as [spin up, spin up, spin down] repeating. For the DMRG calculations, we employed TeNPy's DMRG package with the TwoSiteDMRGEngine, using a two-site update scheme. The maximum bond dimension was set to $\min(N, 300)$, and the energy convergence threshold was set to $10^{-9}$.



\section{Band structure and edge modes in generalized SSH models}

To determine the appropriate filling (or, equivalently, to tune the chemical potential), it is essential to understand the band structure of each model. The two-site SSH model is well-studied and consists of two energy bands\cite{Asb_th_2016}. When the inter-cell hopping amplitude exceeds the intra-cell hopping amplitude, topological edge modes emerge at zero energy, forming a pair of degenerate boundary-localized states. At half-filling, the lower band is fully occupied, placing the Fermi level in the middle of the gap. In this case, the edge modes reside at the Fermi level and become part of the ground state\cite{Asb_th_2016}.

\begin{figure}[H]

\subcaptionbox{\protect\label{a}}
{\includegraphics[width=0.48\linewidth,height=0.42\linewidth]{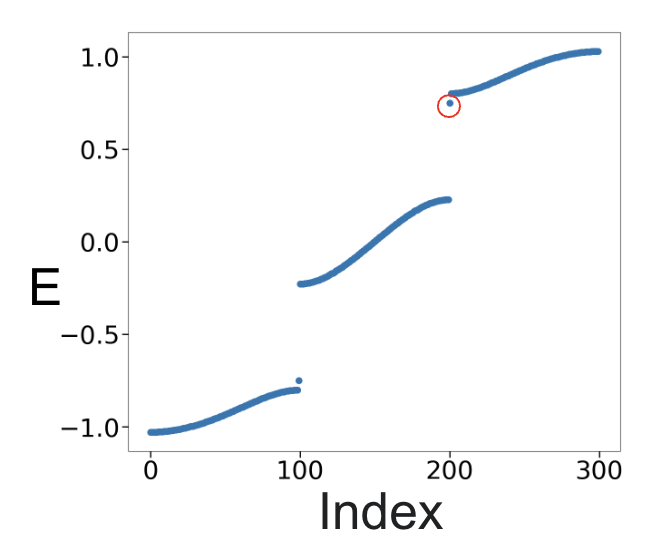}}\hfill
\subcaptionbox{\protect\label{b}}{\includegraphics[width=0.48\linewidth,height=0.42\linewidth]{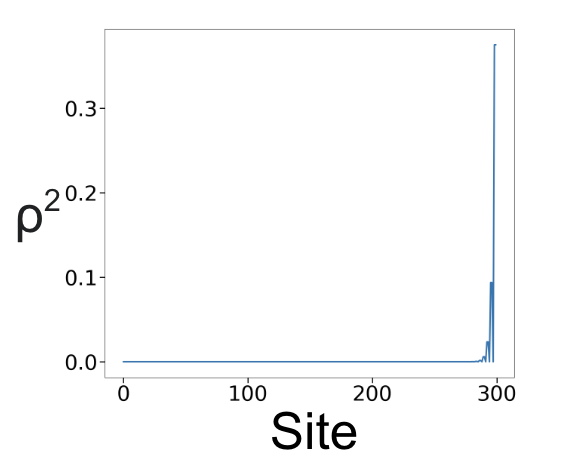}}\hfill
\subcaptionbox{\protect\label{c}}
{\includegraphics[width=0.48\linewidth,height=0.42\linewidth]{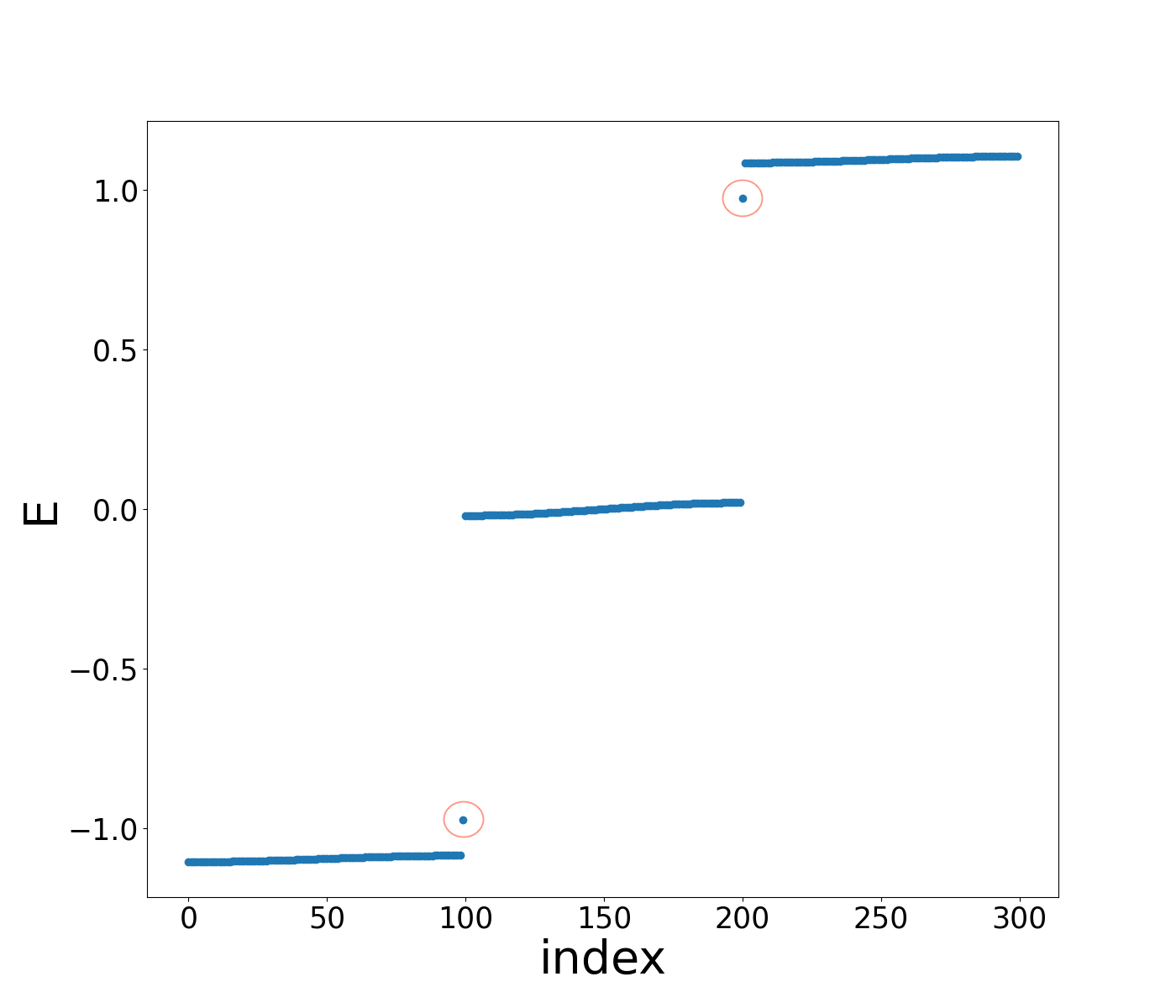}}\hfill
\subcaptionbox{\protect\label{d}}
{\includegraphics[width=0.48\linewidth,height=0.42\linewidth]{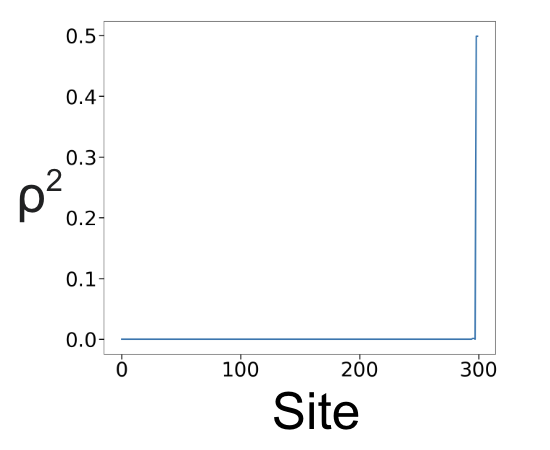}}\hfill
\caption{(a) Energy spectrum of the non-interacting 3-site unit-cell SSH chain with open boundary conditions, N=300, and $\delta$=0.95. The edge modes are highlighted with circles. (b) Wave function amplitude of the edge mode at the bottom of the anti-bonding band with $\delta$=-0.95 (c) Energy spectrum, but at $\delta$=0.20. (d) Wave function amplitude of the edge mode at the bottom of the anti-bonding band for $\delta$=0.20. The localization length of the edge modes is given by
$\xi(\delta) = \left[ \frac{1}{3} \ln \left( \frac{t^2}{t^2 - \delta^2} \right) \right]^{-1}$. Hence it is more strongly localized for larger $\delta$.
}

\label{fig:Band structure}
\end{figure}

In contrast, the three-site SSH model generally supports three bands, bonding, non-bonding, and anti-bonding, and edge modes can emerge either at the top of the bonding band or the bottom of the anti-bonding band\cite{PhysRevB.106.085109,PhysRevA.99.013833,PhysRevB.101.235155}. We have analyzed its band structure and the corresponding edge modes, as shown in Fig. \ref{fig:Band structure}. As the parameter $\delta$ increases, the edge modes become more pronounced. By inspecting their spatial profiles, we find that the lower edge mode consistently appears at site index $N/3 + 1$, and the upper edge mode at $2N/3 + 1$.

Guided by this band structure analysis, we adopt a two-thirds filling configuration—two spin-up states for every spin-down—which ensures that the two lower bands are filled. This allows us to more effectively probe the upper edge mode.

\section{A different cut through the phase diagram: from the topologically trivial phase to the two--edge-state regime in the 3-site unit-cell SSH model}

In the three-site unit-cell SSH model, the non-interacting band structure contains two additional phases beyond the single--edge-state regime: a topologically trivial phase with no edge states and a topologically non-trivial phase hosting two pairs of edge states~\cite{PhysRevB.101.235155,PhysRevB.106.085109}. To explore these phases in the interacting model, we parametrize the couplings in Eq.~(7) as
\begin{align}
    &J^{xy}_1 = (1 + \delta), \quad 
    J^{xy}_2 = \left(1 + \tfrac{1}{2}\delta\right), \quad 
    J^{xy}_3 = 1, \nonumber \\
    &J^z_1 = J_z(1 + \delta), \quad 
    J^z_2 = J_z\left(1 + \tfrac{1}{2}\delta\right), \quad 
    J^z_3 = J_z.
\end{align}
With this parametrization, negative $\delta$ produces two edge-state pairs, while positive $\delta$ yields a trivial phase with no edge states. Because the two edge states are no longer degenerate, DMRG convergence is numerically more stable. As in the single--edge-state analysis, we work at two-thirds filling to ensure that we probe the same many-body sector. Our results are summarized in Figs.~10 and 11: Fig.~10 shows the entanglement entropy (EE) as a function of $\delta$, and Fig.~11 displays the corresponding magnetization profiles.

Analogous to the single--edge-state case, when $J_z > -1/\sqrt{2}$ the system exhibits a single transition at $\delta = 0$. When $J_z < -1/\sqrt{2}$, however, Fig.~10(c) shows that the transition shifts to a negative value of $\delta$, and for $\delta$ values above this transition the system enters an extended region with critical behavior. Only a single transition occurs---rather than two---because in this parametrization $J_1$ and $J_2$ do not vary symmetrically around unity; instead, both increase monotonically with $\delta$. Furthermore, when $J_z < -1$, the system enters a ferromagnetic regime where the EE drops to zero, consistent with the behavior found in the one--edge-state cut.

\begin{figure}[H]

\subcaptionbox{\protect\label{a}}
{\includegraphics[width=0.48\linewidth,height=0.42\linewidth]{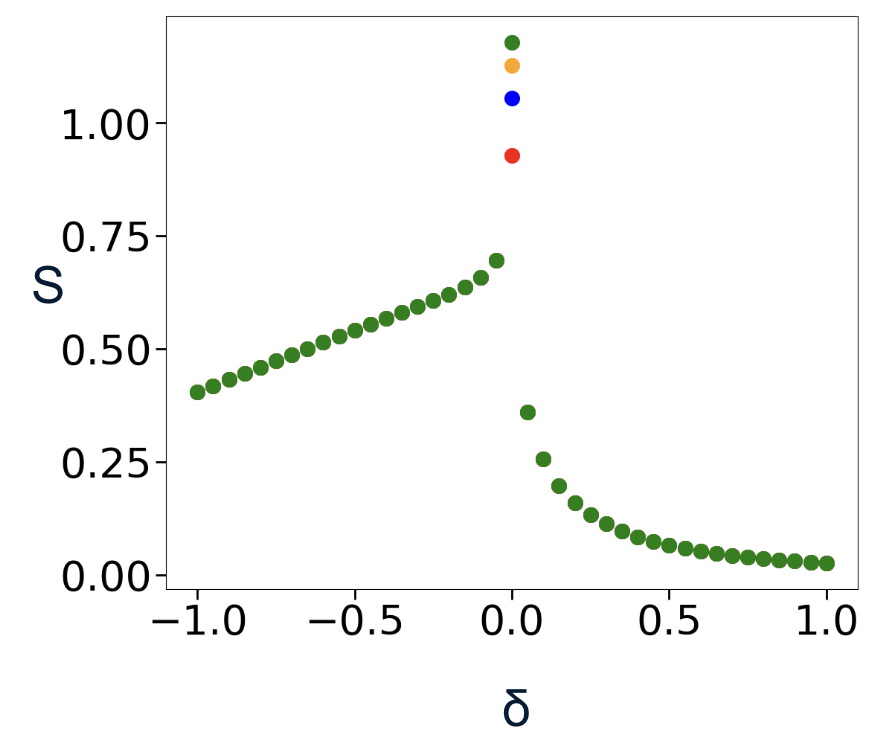}}\hfill
\subcaptionbox{\protect\label{b}}{\includegraphics[width=0.48\linewidth,height=0.42\linewidth]{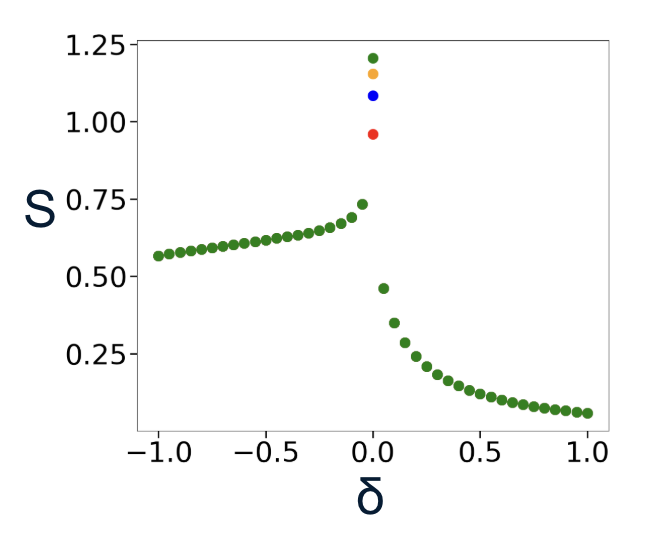}}\hfill
\subcaptionbox{\protect\label{c}}
{\includegraphics[width=0.48\linewidth,height=0.42\linewidth]{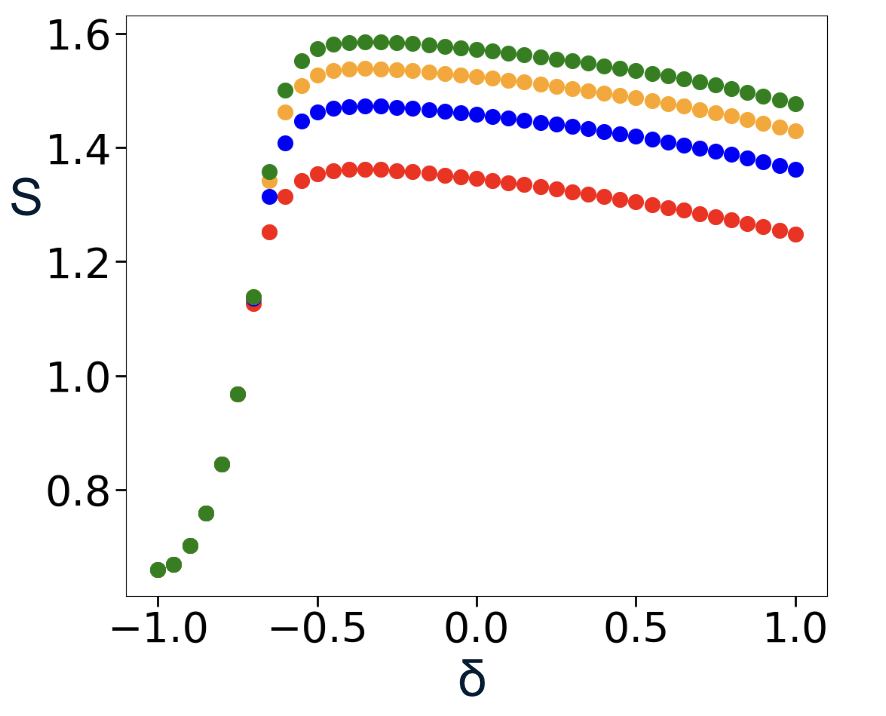}}\hfill
\subcaptionbox{\protect\label{d}}
{\includegraphics[width=0.48\linewidth,height=0.42\linewidth]{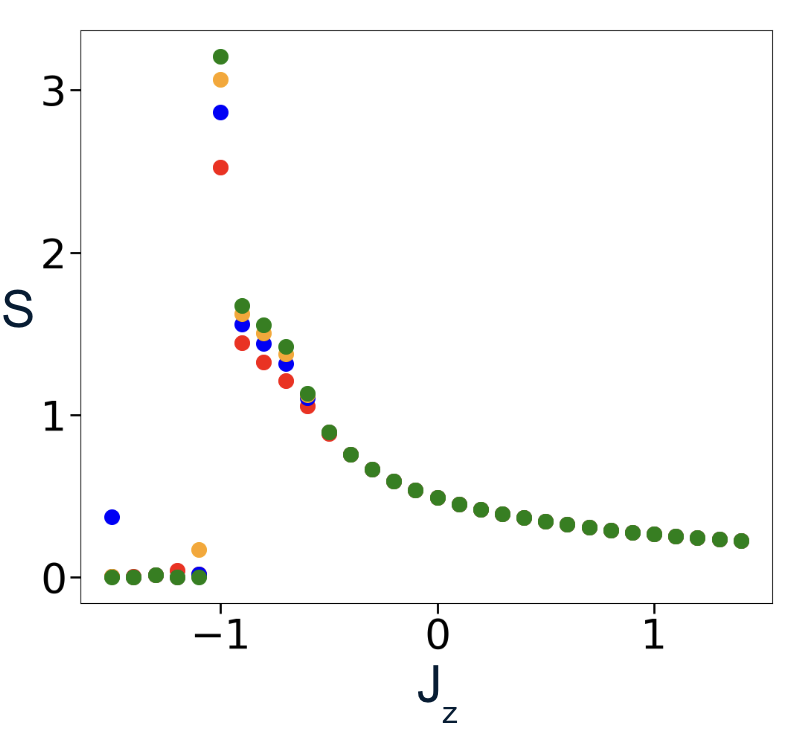}}\hfill
\caption{Bipartite von Neumann entanglement entropy in the interacting 3-site unit cell SSH model at 2/3 filling, evaluated along various cuts in parameter space spanned by (Jz , $\delta$). Here we consider $\delta$ sweeps at fixed interaction strengths (a) Jz =4.0 (b) Jz =2.0 (c) Jz =-0.8. In (d) we fix $\delta$=0.30 and sweep through Jz . Each color represents one fixed length(Red: L=30, Blue: L=60, Orange: L=90, Green: L=120}
\end{figure}

The magnetization profiles (Fig.~11) confirm the persistence of boundary localization. For negative $\delta$, a well-defined surface state appears at the left edge. For positive $\delta$, no edge state forms, and the on-site magnetizations instead organize into a nearly uniform ``stronger--weaker'' pattern throughout the bulk. As in the one--edge-state case, the boundary-localized surface state becomes progressively less prominent as $J_z$ increases.
\\
\begin{figure}[H]
    \subcaptionbox{\protect\label{a}} {\includegraphics[width=0.48\linewidth,height=0.42\linewidth]{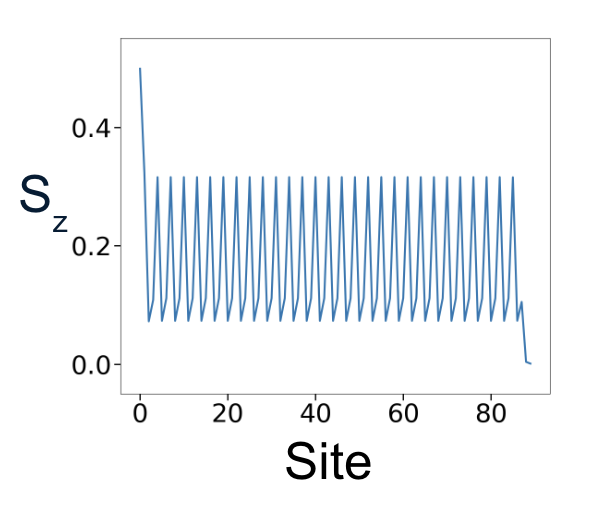}}\hfill
    \subcaptionbox{\protect\label{b}}{\includegraphics[width=0.48\linewidth,height=0.42\linewidth]{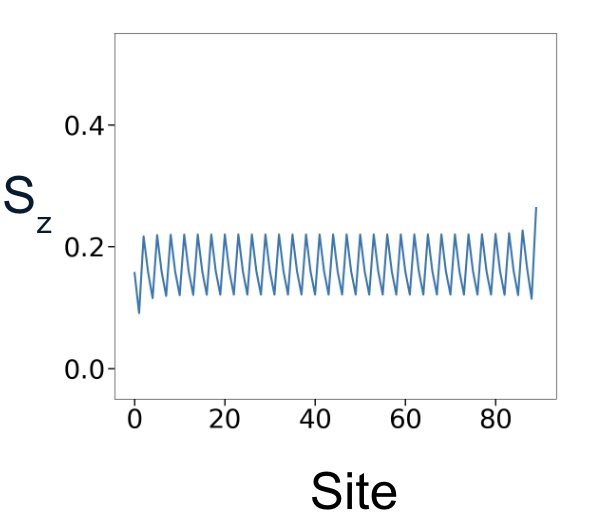}}\hfill
    \caption{Magnetization profiles at (a) $J_z$=-0.5 and $\delta$=-0.95 (b)$J_z$=-0.5 and $\delta$=0.95}
    \label{fig:Magnetization SSH3 nontrivial}
\end{figure}

\end{document}